\documentclass[twocolumn,showkeys]{revtex4}
\usepackage[utf8]{inputenc}
\usepackage{graphicx}

\begin{document}

\title{Waves propagating parallel to the magnetic field in relativistically hot plasmas:
a hydrodynamic model with the average reverse gamma factor evolution}

\author{Pavel A. Andreev}
\email{andreevpa@physics.msu.ru}
\affiliation{Department of General Physics, Faculty of physics, Lomonosov Moscow State University, Moscow, Russian Federation, 119991.}

\date{\today}

\begin{abstract}
The high-frequency part of spectrum of electromagnetic waves propagating parallel to the external magnetic field is considered
for the macroscopically motionless plasmas with the relativistic temperatures $T\sim m_{e}c^{2}$,
where $m_{e}$ is the mass of electron,
$c$ is the speed of light.
The analysis is based on the novel hydrodynamic model
based on four equations for the material fields
which can be combined in two four vectors.
These material fields are the concentration and the velocity field
\emph{and} the average reverse relativistic $\gamma$ functor and the flux of the reverse relativistic $\gamma$ functor.
In the nonrelativistic regime we have three waves (the ions are assumed to be motionless).
Strong thermal effects lead to a coefficient in front of cyclotron frequency
which decreases the effective contribution of the cyclotron frequency.
At $T=0.1m_{e}c^{2}$ we have a decrease of area of existence of fast magneto-sound wave from the area of the large frequencies.
While the area of existence of extraordinary waves becomes larger towards smaller frequencies.
The strong magnetic field limit $\mid\Omega_{e}\mid > \omega_{Le}$ additional wave appears with frequency below thermally decreased cyclotron frequency,
where $\mid\Omega_{e}\mid$ is the electron cyclotron frequency, and $\omega_{Le}$ is the Langmuir frequency.
Further increase of temperature leads to the disappearance of fast magneto-sound wave
and to the considerable increase of area of existence of extraordinary towards smaller frequencies.
\end{abstract}

%\pacs{}% PACS, the Physics and Astronomy
                             % Classification Scheme.

\maketitle

%52.30.Ex	Two-fluid and multi-fluid plasmas
%52.35.Dm	Sound waves

%52.27.Ep	Electron-positron plasmas

%%%%%%%%%%TEXT

%\mbox{\boldmath $\sigma$}

\section{Introduction}

\begin{figure}[h!] \includegraphics[width=8cm,angle=0]{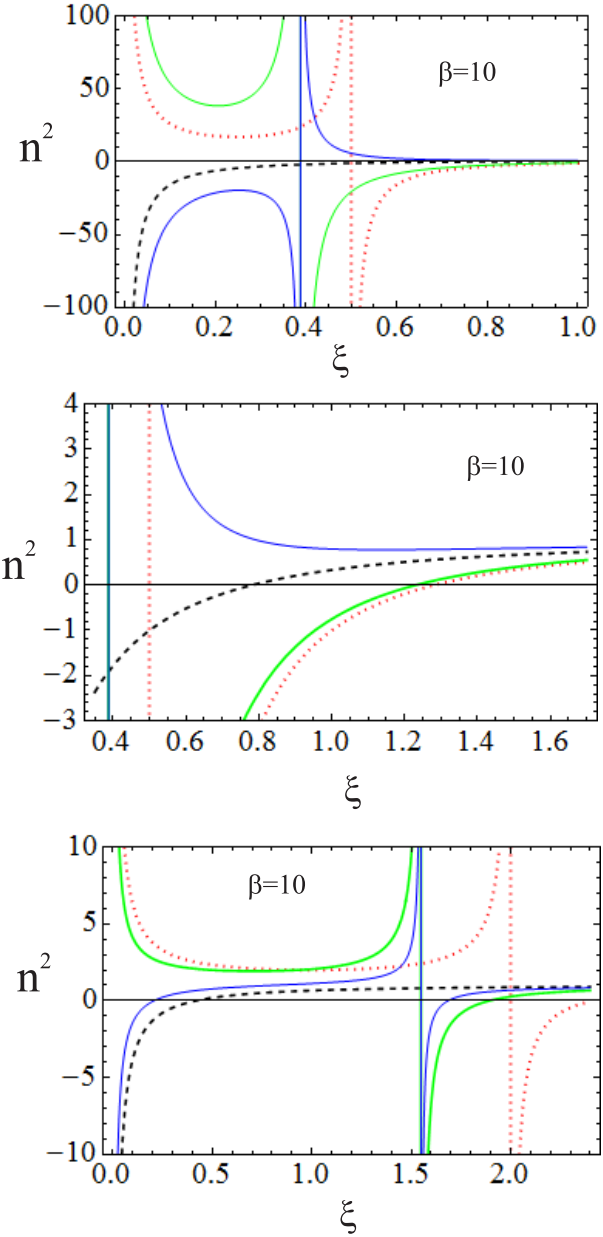}
\caption{\label{RHD2021ClLM Fig 01}
The square of the refractive index square $n^{2}$
as the function of the frequency $n^{2}(\omega)$ is demonstrated to understand the spectrum of the electromagnetic waves
in magnetized and relativistically hot plasmas.
In this figure the relativistic effects are relatively small, so $\beta=mc^{2}/T=10$.
The refractive index square $n^{2}$ is demonstrated in compare with the nonrelativistic regime (it is presented within the dashed lines).
The upper figure is made for small cyclotron frequency $b=\mid\Omega_{e}\mid/\omega_{Le}=0.5$.
The middle figure is made in the same regime, but it shows the high-frequency part of spectrum in more details.
The lower figure is made for the large cyclotron frequency $b=\mid\Omega_{e}\mid/\omega_{Le}=2$.
}
\end{figure}

\begin{figure} \includegraphics[width=8cm,angle=0]{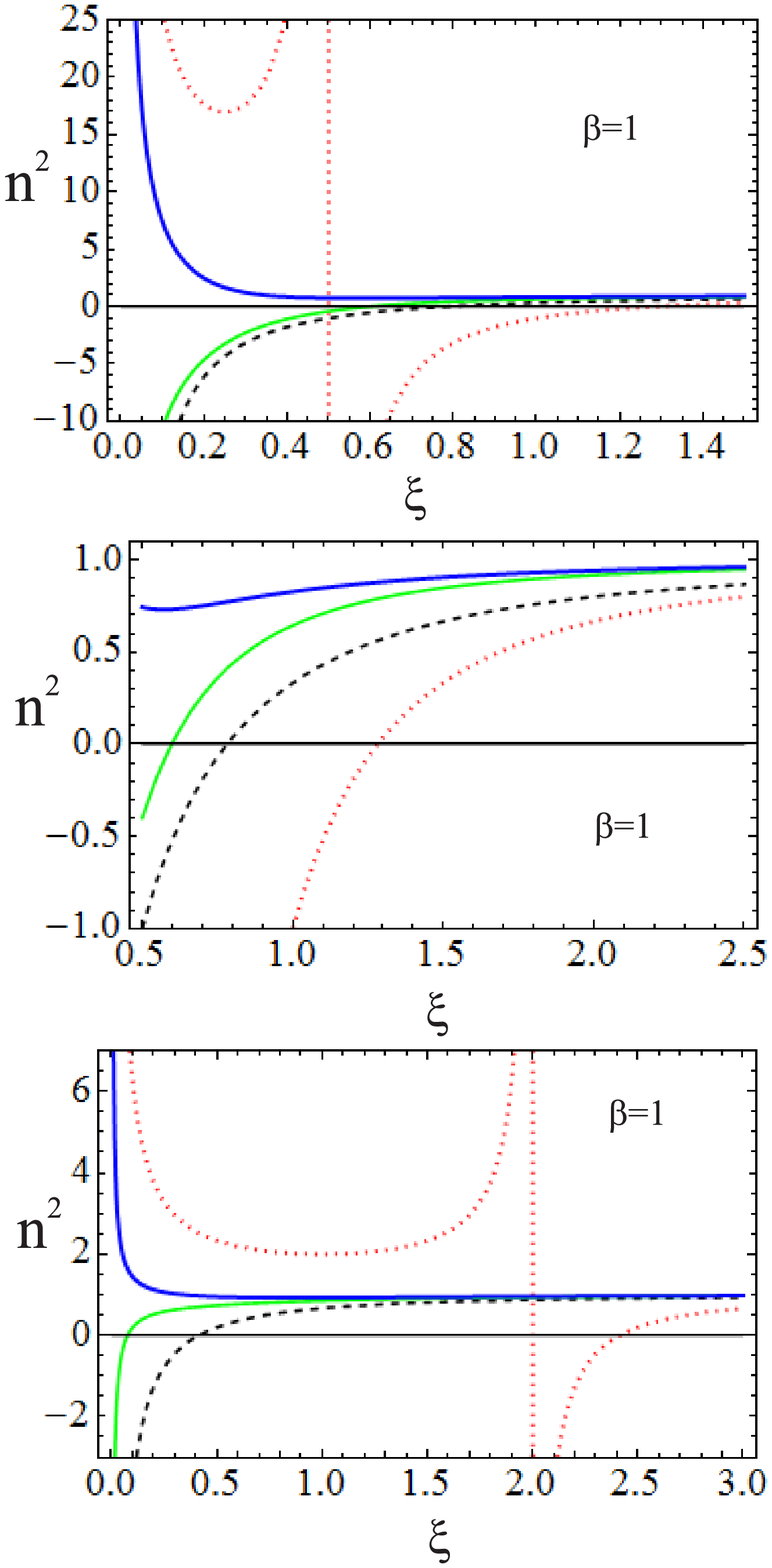}
\caption{\label{RHD2021ClLM Fig 02}
The spectrum similar to figure (\ref{RHD2021ClLM Fig 01}) is demonstrated
for the temperature equal to the rest energy of electron
$\beta=mc^{2}/T=1$.
The upper figure is made for small cyclotron frequency $b=\mid\Omega_{e}\mid/\omega_{Le}=0.5$.
The middle figure is made in the same regime, but it shows the high-frequency part of spectrum in more details.
The lower figure is made for the large cyclotron frequency $b=\mid\Omega_{e}\mid/\omega_{Le}=2$.
}
\end{figure}

\begin{figure} \includegraphics[width=8cm,angle=0]{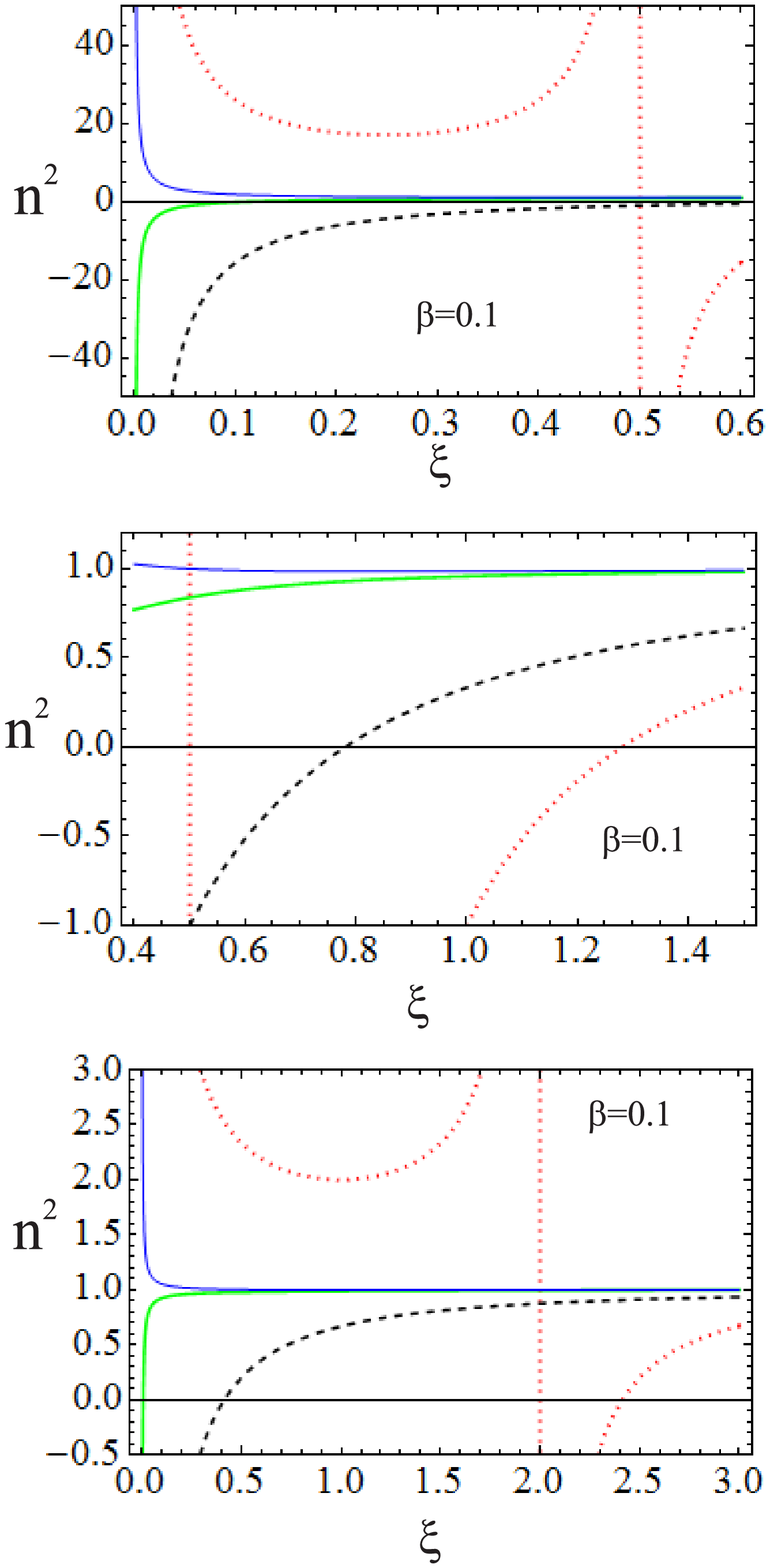}
\caption{\label{RHD2021ClLM Fig 03}
The spectrum similar to figure (\ref{RHD2021ClLM Fig 01}) is demonstrated
for the temperature above the rest energy of electron
$\beta=mc^{2}/T=0.1$.
The upper figure is made for small cyclotron frequency $b=\mid\Omega_{e}\mid/\omega_{Le}=0.5$.
The middle figure is made in the same regime, but it shows the high-frequency part of spectrum in more details.
The lower figure is made for the large cyclotron frequency $b=\mid\Omega_{e}\mid/\omega_{Le}=2$.
}
\end{figure}

Relativistic effects in plasmas like the quantum effects in plasma-like mediums present interesting and important phenomena describing matter in the extreme conditions.
Quantum effects are mostly important for the low temperatures
\cite{Shukla UFN 10}, \cite{Shukla RMP 11}, \cite{Maksimov QHM 99}, \cite{MaksimovTMP 2001}, \cite{Uzdensky RPP 14}.
While the relativistic effects can manifest in different regimes like propagation of relativistic beams in plasmas,
relativistically hot plasmas, and cold high-density plasma, where the temperature Fermi is of order the rest-energy of electron.

Relativistic kinetic model is the most straightforward method of description of collective effects in the relativistically hot plasmas.
However, technically, the kinetic model is rather complex even for nonrelativistic limit.
Therefore, it is necessary to have the hydrodynamic model for relativistically hot plasmas.
There is a widely used model consisting of two equations:
the continuity equation and the Euler equation,
where the Euler equation appears as the equation for the evolution of the four-momentum density $P^{a}$.
Its structure resembles the structure of the nonrelativistic hydrodynamics.
The continuity equation and the Maxwell equations contain the concentration of particles $n$ and the current of particles $\textbf{v}$,
which are combined in the four-velocity vector $v^{a}$ \cite{Shatashvili ASS 97}, \cite{Shatashvili PoP 99}, \cite{Shatashvili PoP 20}:
\begin{equation}\label{RHD2021ClLM cont via v for old model} \partial_{t}n+\nabla\cdot(n\textbf{v})=0,\end{equation}
and
\begin{equation}\label{RHD2021ClLM Euler for P}
\partial_{t}P^{a}+\nabla\cdot (\textbf{v} P^{a})+\nabla^{a} P=
q n\biggl(\textbf{E}+\frac{1}{c}\textbf{v} \times \textbf{B}\biggr),\end{equation}
where $P$ is the pressure.
Hence, the described model requires the additional equation of state to get the relation between the momentum density $P^{a}$ and the four velocity $v^{a}$.
Simple relation can be found for the "cold" plasmas,
where all electrons have same velocity and the relativistic effects are related to the propagation of all electrons as the beam:
$P^{a}=mn \gamma v^{a}$,
where $m$ is the mass of particle, and $\gamma=1/\sqrt{1-v^{2}/c^{2}}$ is the relativistic factor,
with $v^{2}=\textbf{v}^{2}$.
We cannot introduce such relation if there is nonzero temperature,
but it is possible to calculate an approximate equation of state
using the presentation of the momentum density $P^{a}$ and the current of particles $n v^{a}$ via the equilibrium distribution function.
It leads to $P^{a}=(K_{3}(\beta)/K_{2}(\beta))mn \gamma v^{a}$,
where parameter $\beta=mc^{2}/T$ is the reverse dimensionless temperature $T$,
the temperature itself $T$ is presented in the energy units,
and functions $K_{i}(\beta)$ are the $i$-th order Macdonald functions.

Any hydrodynamic and kinetic model requires a truncation.
For instance, the nonrelativistic hydrodynamics of plasmas obtained in the meanfield approximation
includes equation of state for the pressure.
The relativistic hydrodynamic model presented within equations
(\ref{RHD2021ClLM cont via v for old model}) and (\ref{RHD2021ClLM Euler for P})
requires two equations of state.
One for the pressure.
The second equation is for the momentum density,
which is discussed above.
The perturbation of pressure can be found within an extended hydrodynamic model
\cite{Tokatly PRB 99}, \cite{Tokatly PRB 00}, \cite{Miller PoP 16}, \cite{Andreev Ch 21}, \cite{Andreev PoF 21}.
However, some additional functions would appear,
so they would require some equations of state anyway.

It is well known that
the application of the equation of state for the pressure for estimation of the perturbation of pressure gives incorrect coefficients.
For instance, coefficient in corresponding term in the spectrum of Langmuir wave.
In Maxwellian plasmas we have coefficient 1 instead of 3.
For the degenerate plasmas we have coefficient $1/3$ instead of $3/5$.
The application of the extended hydrodynamic model gives systematic corrections of these coefficients \cite{Tokatly PRB 99}, \cite{Tokatly PRB 00}.

One of the major collective plasma phenomena is the Langmuir wave.
Its frequency is approximately equal to the Langmuir frequency $\omega_{Le}=\sqrt{4\pi e^{2}n_{0}/m_{e}}$.
There is increase of frequency with the increase of the wave vector $k$: $\omega=\sqrt{\omega_{Le}^{2}+\alpha k^{2}v_{s}^{2}}$,
and this increase is related to the thermal effects (to the Pauli blocking in the degenerate plasmas),
where coefficient $\alpha$ is discussed above.
Hence, the coefficient in front of the Langmuir frequency, which is the major term, can be obtained from the momentum balance equation
at the application of the first term on the left-hand side and the first term on the right-hand side
(see eq. \ref{RHD2021ClLM Euler for P} for the illustration) .

Some rough estimations in equation of state for the pressure would affect minor term.
But the accuracy of the equation of state for the momentum density would directly affect the major term.
The example of uncertainty of pressure discussed above gives us a hint
that the application of the equation of state for the momentum density based on the equilibrium distribution is not completely reliable.
Thus, we suggest and apply relativistic hydrodynamic model,
where no equation of state is used for the perturbations of major functions.

The Euler equation is the equation of motion of the plasmas,
it demonstrates the change of the four-momentum.
Hence, it is questionable to make the approximate transition via equation of state in the major term in the model.
Therefore, another model is suggested to get description of the plasmas with the relativistically large temperatures \cite{Andreev 2021 05}.
This model is derived from the microscopic motion of the relativistic charged particles
\cite{Kuz'menkov 91}, \cite{Drofa TMP 96}, \cite{Andreev PIERS 2012}, \cite{Kuzmenkov CM 15}.
It consists of four equations.
Derivation starts with the microscopic definition of the concentration.
The evolution of concentration leads to the continuity equation and the definition of the current of particles.
The velocity field is introduced as the ratio of the current of particles to the concentration.
The second equation is the equation of evolution of the current of particles (evolution of the velocity field).
This equation contains for novel functions (in the meanfield approximation).
First, it is obviously symmentric second rank tensor of the flux of the current of particles.
The terms describing interaction contain the density of the reverse relativistic gamma-functor of particles,
the current of the reverse relativistic gamma-functor of particles,
and the second rank tensor of the flux of current of the reverse relativistic gamma-functor of particles.
Equations for the scalar reverse relativistic gamma-functor of particles
and the vector current of the reverse relativistic gamma-functor of particles are obtained.
Their evolution is mainly expressed via the concentration and current of particles.
Hence, the set of equation try to close itself.
However, new functions appear as well
since the large number of the degrees of freedom on the microscopic level manifests via the large number of macroscopic functions.
The evolution of the reverse relativistic gamma-functor of particles leads to already introduced
current of the reverse relativistic gamma-functor of particles, and interaction.
The interaction includes the electric filed only.
The interaction is partially expressed via the product of the velocity field and the electric field.
It also contains a relativistic term
which requires truncation like tensors
the flux of the current of particles
and
the flux of current of the reverse relativistic gamma-functor of particles.
The evolution of current of the reverse relativistic gamma-functor leads to
the flux of current of the reverse relativistic gamma-functor and interaction.
The interaction is expressed via the electromagnetic field, concentration, velocity field, flux of the current of particles,
and the additional function (the fourth rank tensor) requiring the equation of state.
Necessary equations of state are obtained either \cite{Andreev 2021 05}.

Extended sets of hydrodynamic equations allows to calculate equations of state for the hydrodynamic models applying smaller number of functions.
However, to some extend, it is possible to avoid the derivation and the application of the extended hydrodynamic models.
Equilibrium distribution functions can be used to calculate necessary equations of state.
For the relativistic plasmas we can use following distribution function
\begin{equation}\label{RHD2021ClLM distrib func eq}
f_{0}(p)=\textrm{Z} e^{-\epsilon/T},
\end{equation}
where
\begin{equation}\label{RHD2021ClLM normalization of f}
\textrm{Z}=\frac{n}{4\pi m^{2}cTK_{2}(\frac{mc^{2}}{T})},
\end{equation}
with $T$ is the equilibrium temperature in the energy units,
$p$ is the momentum,
$K_{2}(\xi)$ is the second order Macdonald function,
and $\epsilon=\sqrt{m^{2}c^{4}+p^{2}c^{2}}$.
Hence, we use it below to get a number of equations of state.
We consider the magnetized plasmas,
therefore, the application of the isotropic distribution function does not reflect all major physical phenomena.
The phenomena related to the two temperature regimes, one (another) temperature for the motion parallel (perpendicular) to the magnetic field.
However, the isotropic limit can show some essential effects related to the relativistic phenomena.

This paper is organized as follows.
In Sec. II the relativistic hydrodynamic equations are presented and discussed.
In Sec. III the spectrum of collective excitations is considered analytically.
In Sec. IV numerical analysis of obtained spectra is demonstrated.
In Sec. V a brief summary of obtained results is presented.

%{\color{blue}}

\section{Relativistic hydrodynamic model}
%{\color{blue}}

Here we follow Ref. \cite{Andreev 2021 05},
where the following set of hydrodynamic equations
is derived for the relativistic plasmas with the relativistic temperature.
It consists of four equations (in three-vector notations).
First equation is the continuity equation
\begin{equation}\label{RHD2021ClLM cont via v} \partial_{t}n+\nabla\cdot(n\textbf{v})=0.\end{equation}
Next, the velocity field evolution equation is
$$n\partial_{t}v^{a}+n(\textbf{v}\cdot\nabla)v^{a}+\frac{\partial^{a}p}{m}
=\frac{e}{m}\Gamma E^{a}+\frac{e}{mc}\varepsilon^{abc}(\Gamma v_{b}+t_{b})B_{c}$$
\begin{equation}\label{RHD2021ClLM Euler for v} -\frac{e}{mc^{2}}(\Gamma v^{a} v^{b}+v^{a}t^{b}+v^{b}t^{a})E_{b}
-\frac{e}{mc^{2}}\tilde{t}E^{a}. \end{equation}
It includes the flux of the thermal velocities $p$
which is similar to the pressure,
but the pressure is the flux of momentum.
The interaction of electrons via the electromagnetic field is presented within four terms placed on the right-hand side.
It is found in the mean-field approximation (the self-consistent field approximation).
Parameters $m$ and $e$ are the mass and charge of particle,
$c$ is the speed of light,
$\delta^{ab}$ is the three-dimensional Kronecker symbol,
$\varepsilon^{abc}$ is the three-dimensional Levi-Civita symbol.
In equation (\ref{RHD2021ClLM Euler for v}) and below we assume the summation on the repeating index
$v^{b}_{s}E_{b}=\sum_{b=x,y,z}v^{b}_{s}E_{b}$.
Moreover, the metric tensor has diagonal form corresponding to the Minkovskii space,
it has the following sings $g^{\alpha\beta}=\{-1, +1, +1, +1\}$.
Hence, we can change covariant and contrvariant indexes for the three-vector indexes: $v^{b}_{s}=v_{b,s}$.
The Latin indexes like $a$, $b$, $c$ etc describe the three-vectors,
while the Greek indexes are deposited for the four-vector notations.
The Latin indexes can refer to the species $s=e$ for electrons or $s=i$ for ions.
The Latin indexes can refer to the number of particle $j$ at the microscopic description.
However, the indexes related to coordinates are chosen from the beginning of the alphabet,
while other indexes are chosen in accordance with their physical meaning.

The equation of evolution of the averaged reverse relativistic gamma factor, called here the hydrodynamic Gamma function, is
\begin{equation}\label{RHD2021ClLM eq for Gamma} \partial_{t}\Gamma+\partial_{b}(\Gamma v^{b}+t^{b})
=-\frac{e}{mc^{2}}n\textbf{v}\cdot\textbf{E}\biggl(1-\frac{1}{c^{2}}\biggl(\textbf{v}^{2}+\frac{5p}{n}\biggr)\biggr).\end{equation}
This function appears on the right-hand side of the velocity field evolution equation together with two other functions.
They are the vector of current of the reverse relativistic gamma factor
and the second rank tensor describing the flux of current of the reverse relativistic gamma factor.

The final equation in this set of hydrodynamic equations is the equation of evolution of
current of the reverse relativistic gamma factor (the hydrodynamic Theta function).
Actually it is equation for its thermal part $t^{a}$:
$$(\partial_{t}+\textbf{v}\cdot\nabla)t^{a}+\partial^{a}\tilde{t}
+(\textbf{t}\cdot\nabla) v^{a}+t^{a} (\nabla\cdot \textbf{v})$$
$$+\Gamma(\partial_{t}+\textbf{v}\cdot\nabla)v^{a}
=\frac{e}{m}nE^{a}\biggl[1-\frac{\textbf{v}^{2}}{c^{2}}-\frac{3p}{nc^{2}}\biggr]$$
$$+\frac{e}{mc}\varepsilon^{abc}nv_{b}B_{c}\biggl[1-\frac{\textbf{v}^{2}}{c^{2}}-\frac{5p}{nc^{2}}\biggr]
-\frac{2e}{mc^{2}}E^{a}p\biggl[1-\frac{\textbf{v}^{2}}{c^{2}}\biggr]$$
\begin{equation}\label{RHD2021ClLM eq for t a} -\frac{e}{mc^{2}}nv^{a}v^{b}E_{b}\biggl[1-\frac{\textbf{v}^{2}}{c^{2}}-\frac{9p}{nc^{2}}\biggr]
+\frac{10e}{3mc^{4}}M E^{a}.\end{equation}
We use equation of state for the second rank tensor describing the flux of current of the reverse relativistic gamma factor,
which enters equation (\ref{RHD2021ClLM eq for t a}).

The equations for the material fields (\ref{RHD2021ClLM cont via v})-(\ref{RHD2021ClLM eq for t a})
are coupled to the Maxwell equations
$ \nabla \cdot\textbf{B}=0$,
\begin{equation}\label{RHD2021ClLM rot E and div E} \begin{array}{cc}
\nabla\times \textbf{E}=-\frac{1}{c}\partial_{t}\textbf{B}, & \nabla \cdot\textbf{E}=4\pi(en_{i}-en_{e}),
\end{array}
\end{equation}
and
\begin{equation}\label{RHD2021ClLM rot B with time}
\nabla\times \textbf{B}=\frac{1}{c}\partial_{t}\textbf{E}+\frac{4\pi q_{e}}{c}n_{e}\textbf{v}_{e},\end{equation}
where the ions are considered as the motionless positively charged background.

%\begin{equation}\label{RHD2021ClLM div B} \nabla \cdot\textbf{B}=0,\end{equation}
%\begin{equation}\label{RHD2021ClLM rot E} \nabla\times \textbf{E}=-\frac{1}{c}\partial_{t}\textbf{B},\end{equation}
%\begin{equation}\label{RHD2021ClLM div E with time} \nabla \cdot\textbf{E}=4\pi(en_{i}-en_{e}),\end{equation}

\section{Electromagnetic waves in the relativistic magnetized plasmas}

The small amplitude waves are considered.
The macroscopically motionless plasmas is considered in the equilibrium state.
Its equilibrium state can be described by the relativistic Maxwellian distribution.
Hence, required equilibrium equations of state can be gained from this distribution.
The equilibrium concentration $n_{0}$ is nonzero.
The equilibrium velocity field $\textbf{v}_{0}$ and the equilibrium electric field $\textbf{E}_{0}$ are equal to zero.
The external magnetic field is constant and uniform.
It is directed along $Oz$ axis $\textbf{B}_{0}=B_{0}\textbf{e}_{z}$.
The pressure tensor $p^{ab}$ and tensor $t^{ab}$ are assumed to be diagonal tensors:
$p^{ab}=p\delta^{ab}$ and $t^{ab}=\tilde{t}\delta^{ab}$.
The "diagonal" form is assumed for tensor $M^{abcd}$ as well:
$M^{abcd}=M_{0}(\delta^{ab}\delta^{cd}+\delta^{ac}\delta^{bd}+\delta^{ad}\delta^{bc})/3$.
Functions $\Gamma_{0}$, $\textbf{t}_{0}$, $p_{0}$, $\tilde{t}_{0}$, $\textbf{q}_{0}$, $M_{0}$ describe the equilibrium state.
The perturbations $\delta p$, $\delta \tilde{t}$ require some equations of state.
The equilibrium expressions for functions
$p$, $\tilde{t}$, $\textbf{q}$, $M$ are used as the equations of state for the nonequilibrium functions.
Approximate calculation of functions $p^{ab}$, $t^{ab}$, $\textbf{q}$, and $M^{abcd}$ gives to the following representations
$p^{ab}=c^{2}\delta^{ab}\tilde{Z}f_{1}(\beta)/3$,
$t^{ab}=c^{2}\delta^{ab}\tilde{Z}f_{2}(\beta)/3$
$M^{abcd}=c^{4}(\delta^{ab}\delta^{cd}+\delta^{ac}\delta^{bd}+\delta^{ad}\delta^{bc})\tilde{Z}f_{3}(\beta)/15$,
and
$\textbf{q}=0$,
where
$\beta=mc^{2}/T$,
$\tilde{Z}=4\pi Z (mc)^{3}=n\beta K_{2}^{-1}(\beta)$,
\begin{equation}\label{RHD2021ClLM f 1} f_{1}(\beta)=\int_{1}^{+\infty}\frac{d x}{x}(x^{2}-1)^{3/2}e^{-\beta x}, \end{equation}
\begin{equation}\label{RHD2021ClLM f 2} f_{2}(\beta)=\int_{1}^{+\infty}\frac{d x}{x^{2}}(x^{2}-1)^{3/2}e^{-\beta x}, \end{equation}
and
\begin{equation}\label{RHD2021ClLM f 3} f_{3}(\beta)=\int_{1}^{+\infty}\frac{d x}{x^{3}}(x^{2}-1)^{5/2}e^{-\beta x}. \end{equation}
Functions $f_{1}(\beta)$, $f_{2}(\beta)$ and $f_{3}(\beta)$ are calculated numerically below for the chosen values of temperatures.
For each function describing the thermal we introduce corresponding velocity
$\delta p=U_{p}^{2} \delta n$,
$\delta \tilde{t}=U_{t}^{2} \delta n$,
$\delta M=U_{M}^{4} \delta n$.

Let us present the linearized equations.
The linearized equations for the evolution of the three projections of velocity field are
\begin{equation}\label{RHD2021ClLM velocity field evolution equation lin 1D x}
n_{0}\partial_{t}\delta v_{x}
=\frac{e}{m}\Gamma_{0} \delta E_{x}-\frac{e}{mc^{2}}\tilde{t}_{0}\delta E^{x}+\Omega_{e}(\Gamma_{0}\delta v_{y}+\delta t_{y}),
\end{equation}
\begin{equation}\label{RHD2021ClLM velocity field evolution equation lin 1D y}
n_{0}\partial_{t}\delta v_{y}
=\frac{e}{m}\Gamma_{0} \delta E_{y}-\frac{e}{mc^{2}}\tilde{t}_{0}\delta E^{y}-\Omega_{e}(\Gamma_{0}\delta v_{x}+\delta t_{x}),
\end{equation}
\begin{equation}\label{RHD2021ClLM velocity field evolution equation lin 1D z}
n_{0}\partial_{t}\delta v_{z}+\partial_{z}\delta p
=\frac{e}{m}\Gamma_{0} \delta E_{z}-\frac{e}{mc^{2}}\tilde{t}_{0}\delta E^{z}.
\end{equation}

The Maxwell equations lead to
\begin{equation}\label{RHD2021ClLM Maxwell lin wave}
(\omega^{2}-k_{z}^{2}c^{2})\delta \textbf{E}+4\pi q_{e}\imath\omega n_{0}\delta \textbf{v}=0. \end{equation}

The z-projection of the velocity field together with the z-projection of the electric field and the concentration
satisfying linearized continuity equation (\ref{RHD2021ClLM cont via v}):
\begin{equation}\label{RHD2021ClLM continuity equation lin 1D}
\partial_{t}\delta n+n_{0}\partial_{z} \delta v_{z}=0 \end{equation}
describe the longitudinal waves described in Ref. \cite{Andreev 2021 05}

%the Poisson equation
%\begin{equation}\label{RHD2021ClLM Poisson equation lin}
%\partial_{x}\delta E_{x}=-4\pi \mid e\mid \delta n. \end{equation}

The dynamics of the x- and y-projections of the velocity and the x- and y-projections of the electric field require
the equations for evolution of the x- and y-projections of the flux of the reverse relativistic factor:
$$\partial_{t}\delta t_{x} +\Gamma_{0} \partial_{t}\delta v_{x}$$
\begin{equation}\label{RHD2021ClLM evolution of Theta lin 1D x}
=n_{0}\biggl( 1-\frac{5 p_{0}}{n_{0}c^{2}}\biggr)\biggl[\frac{e}{m}\delta E_{x}+\Omega_{e}\delta v_{y}\biggr] +\frac{10e}{3mc^{4}}M_{0}\delta E_{x}, \end{equation}
and
$$\partial_{t}\delta t_{y} +\Gamma_{0} \partial_{t}\delta v_{y}$$
\begin{equation}\label{RHD2021ClLM evolution of Theta lin 1D y}
=n_{0}\biggl( 1-\frac{5 p_{0}}{n_{0}c^{2}}\biggr)\biggl[\frac{e}{m}\delta E_{y}-\Omega_{e}\delta v_{x}\biggr] +\frac{10e}{3mc^{4}}M_{0}\delta E_{y}. \end{equation}
where $M_{0}^{xxcc}=(5/3)M_{0}$.

The evolution of the average reverse relativistic factor is not involved in the equations presented above.
Its equation
\begin{equation}\label{RHD2021ClLM evolution of Gamma lin 1D}
\partial_{t}\delta\Gamma +\Gamma_{0}\partial_{z}\delta v_{z}+\partial_{z}\delta t_{z} =0, \end{equation}
together with evolution of $\delta t_{z}$
do not give any additional solutions either.

To consider various temperature regimes we choose three regimes $T_{1}=0.1 mc^{2}$, $T_{2}=mc^{2}$, and $T_{3}=10mc^{2}$.
It leads to the following values of the dimensionless parameter $\beta=mc^{2}/T$:
$\beta_{1}=10$, $\beta_{2}=1$, and $\beta_{3}=0.1$.
The following values of parameters defining the equilibrium values of parameters and equations of state in these regimes are calculated.
For $\beta_{2}=1$, we obtain $K_{1}/K_{2}=0.38$, $U_{t}^{2}/c^{2}=\beta f_{2}/(3 K_{2})=0.1$,
$U_{p}^{2}/c^{2}=\beta f_{1}/(3 K_{2})=0.28$, and $U_{M}^{4}/c^{4}=\beta f_{3}/(5 K_{2})=0.15$,
where $K_{1}(1)=0.6$, $K_{2}(1)=1.6$, $f_{1}(1)=1.35$, $f_{2}(1)=0.46$, $f_{3}(1)=1.17$.
Next, for $\beta_{3}=0.1$, we find $K_{1}/K_{2}=0.05$, $U_{t}^{2}/c^{2}=\beta f_{2}/(3 K_{2})=0.02$,
$U_{p}^{2}/c^{2}=0.33$, and $U_{M}^{4}/c^{4}=0.2$,
where $K_{1}(0.1)=10$, $K_{2}(0.1)=200$, $f_{1}(0.1)=2\times 10^{3}$, $f_{2}(0.1)=100$, $f_{3}(0.1)=2\times 10^{3}$.
Finally, for $\beta_{1}=10$,
we have $K_{1}/K_{2}=0.91$, $U_{t}^{2}/c^{2}=0.07$,
$U_{p}^{2}/c^{2}=0.08$, and $U_{M}^{4}/c^{4}=0.02$,
where
$K_{1}(10)=2\times 10^{-5}$, $K_{2}(10)=2.2\times10^{-5}$, $f_{1}(10)=5\times 10^{-7}$,
$f_{2}(10)=4.2\times10^{-7}$, $f_{3}(10)=1.7\times 10^{-7}$.

We start the discussion of the spectrum with the representation of the spectrum in nonrelativistic regime
found from the standard hydrodynamic model based on the continuity and Euler equations:
\begin{equation}\label{RHD2021ClLM Spectrum nonRel}
\omega^{2}-k^{2}c^{2}-\omega_{Le}^{2}\frac{\omega}{\omega\mp\Omega_{e}}=0,
\end{equation}
where
$\omega_{Le}^{2}=4\pi e^{2}n_{0}/m$ is the Langmuir frequency,
$\Omega_{e}=-\mid \Omega_{e}\mid=q_{e}B_{0}/mc$.

Next, we present corresponding spectrum found from equations
(\ref{RHD2021ClLM velocity field evolution equation lin 1D x}), (\ref{RHD2021ClLM velocity field evolution equation lin 1D y}),
(\ref{RHD2021ClLM Maxwell lin wave}),
(\ref{RHD2021ClLM evolution of Theta lin 1D x}), and (\ref{RHD2021ClLM evolution of Theta lin 1D y}):
$$\omega^{2}-k^{2}c^{2}-\frac{\omega_{Le}^{2}}{\omega^{2}-\Omega_{e}^{2}(1-5\frac{u_{p}^{2}}{c^{2}})}\times$$
\begin{equation}\label{RHD2021ClLM Spectrum}
\times\biggl[\omega^{2}\biggl(\frac{\Gamma_{0}}{n_{0}}-\frac{u_{t}^{2}}{c^{2}}\biggr) \pm\omega\Omega_{e}\biggl(1-5\frac{u_{p}^{2}}{c^{2}}+\frac{10u_{M}^{4}}{3c^{4}}\biggr)\biggr]=0.
\end{equation}
If the temperature is rather small
we have that
the characteristic velocities $u_{p}$, $u_{t}$, $u_{M}$ are small in compare with the speed of light.
So, we have $u_{p}\ll c$, $u_{t}\ll c$, $u_{M}\ll c$ and $\Gamma_{0}\rightarrow n_{0}$.
Hence, the thermal effects can be neglected in (\ref{RHD2021ClLM Spectrum}),
appearing ratio $(\omega\pm\Omega_{e})/(\omega^{2}-\Omega_{e}^{2})$ simplifies to $1/(\omega\mp\Omega_{e})$,
and we obtain equation (\ref{RHD2021ClLM Spectrum nonRel}).
We have $1/(\omega\mp\Omega_{e})=1/(\omega\pm\mid\Omega_{e}\mid)$ in equation (\ref{RHD2021ClLM Spectrum nonRel}).
The upper sign corresponds to the slower extraordinary wave.
The lower sign leads to the pole in this fraction at $\omega=\mid\Omega_{e}\mid$.
This frequency divides range of frequencies on two areas.
Each of them contains the wave solution.
Fast magneto-sound wave in the small frequency area
and the fast extraordinary wave in the large frequency area.

Presence of the temperature effects does not allow to cancel parts of numerator and denominator in the last term in equation (\ref{RHD2021ClLM Spectrum}).
Hence, both branches of the spectrum have pole at $\omega=\mid\Omega_{e}\mid \sqrt{1-5u_{p}^{2}/c^{2}}$.
While this pole is shifted from the cyclotron frequency in the area of smaller frequencies.

We present the analysis of spectrum based on the refractive index.
Hence, we start with the representation of equation (\ref{RHD2021ClLM Spectrum nonRel})
via corresponding expression for the refractive index:
\begin{equation}\label{RHD2021ClLM Index nonRel}
n^{2}=1-\frac{1}{\xi(\xi\pm b)},
\end{equation}
where we use the definition of the refractive index $n=kc/\omega$ and dimensionless cyclotron frequency $b=\mid\Omega_{e}\mid/_{Le}$.

Let us represent the dispersion equation (\ref{RHD2021ClLM Spectrum}) as the expression for the refractive index
which gives the generalization of equation (\ref{RHD2021ClLM Index nonRel}):
\begin{equation}\label{RHD2021ClLM INDEX}
n^{2}=1-\frac{[\xi(g-\varsigma^{2})\pm b(1-5p^{2}+10 M^{4}/3)]}{\xi(\xi^{2}-b^{2}(1-5p^{2}))},
\end{equation}
where $g=\Gamma_{0}/n_{0}$, $\varsigma=u_{t}/c$, $p=u_{p}/c$, and $M=u_{M}/c$.
Physical effects entering equation (\ref{RHD2021ClLM INDEX}) reflects all described after equation (\ref{RHD2021ClLM Spectrum}).
So we do nor repeat this discussion.

\subsection{Analysis of the spectrum}

Let us remind the properties of the electromagnetic waves propagation parallel to the magnetic field
in the limit of small nonrelativistic temperatures and motionless ions.
We have three wave solutions: the fast magneto-sound wave, the slow extraordinary wave, and the fast extraordinary wave.
Some times notion "extraordinary wave" is reserved for the waves propagating perpendicular to the magnetic field,
but we use it for the high-frequency circularly polarized waves propagating parallel to the magnetic field either.
Frequency of the fast extraordinary wave (the slow extraordinary wave) is above of
$\omega_{0}^{(1)}=\sqrt{\omega_{Le}^{2}+\Omega_{e}^{2}/4}+\mid\Omega_{e}\mid/2$
(of $\omega_{0}^{(3)}=\sqrt{\omega_{Le}^{2}+\Omega_{e}^{2}/4}-\mid\Omega_{e}\mid/2$),
it grows steady with the increase of the wave vector.
For its refractive index
we find that
it goes to $-\infty$ at $\omega\rightarrow +\mid\Omega_{e}\mid$ (at $\omega\rightarrow +0$).
It grows up to zero value $n\rightarrow0$ at $\omega_{0}^{(1)}$ (at $\omega_{0}^{(3)}$).
The wave exists at $n^{2}>0$.
So, the wave exists at $\omega >\omega_{0}^{(1)}$ (at $\omega >\omega_{0}^{(3)}$).
Corresponding $n^{2}$ grows from $0$ to $1$ at the increase of $\omega$ from $\omega_{0}^{(1)}$ (from $\omega_{0}^{(3)}$) up to $+\infty$.

The third wave is the fast magneto-sound wave.
Its frequency is restricted by $\omega\in (0,\mid\Omega_{e}\mid)$.
The square of the refractive index goes to infinity at $\omega\rightarrow0$ and at $\omega\rightarrow\mid\Omega_{e}\mid-0$.

Next, we consider the spectrum at the relatively small relativistic temperatures $\beta=10$.
We start with the discussion of the small magnetic field regime $b=\mid\Omega_{e}\mid/\omega_{Le}=0.5$.
It is presented in the upper and middle figures in Fig. \ref{RHD2021ClLM Fig 01}.
The asymptotic of fast extraordinary wave is shifted in area of smaller frequencies
from $\mid\Omega_{e}\mid$ to $\mid\Omega_{e}\mid \sqrt{1-5p^{2}}$.
Hence, the minimal frequency of the fast extraordinary wave is shifted in area of smaller frequencies as well $\omega_{min,FEx}<\omega_{0}^{(1)}$,
see points of zero refractive index in middle figure in Fig. \ref{RHD2021ClLM Fig 01}
for the green continuous $\omega_{min,FEx}$ and red dashed $\omega_{0}^{(1)}$ lines.
Dependence of $n^{2}(\omega)$ has larger modification for the slower extraordinary wave.
Major change is in the increase of area of existence of wave.
Its $\omega_{min,SEx}$ decreases from $\omega_{0}^{(3)}$ to $\mid\Omega_{e}\mid \sqrt{1-5p^{2}}$.
However, the area of existence of the fast magneto-sound wave decreases down to $\omega\in (0, \mid\Omega_{e}\mid \sqrt{1-5p^{2}})$.

Let us describe the spectrum in the same temperature regime $\beta=10$,
but for the large magnetic field $b=\mid\Omega_{e}\mid/\omega_{Le}=2$.
It is shown in lower figure in Fig. \ref{RHD2021ClLM Fig 01}.
The fast extraordinary wave and the fast magneto-sound wave show same behavior as it is described for the small magnetic field.
However, the minimal frequency of the fast extraordinary wave has relatively large shift in area of smaller frequencies $\omega_{min,FEx}\ll \omega_{0}^{(1)}$.
Moreover, we find $\omega_{min,FEx}<\mid\Omega_{e}\mid$.
However, the slow extraordinary wave demonstrates an interesting behavior.
Its splits on two wave solutions.
The high frequency part exists in the area above thermally shifted cyclotron frequency $\mid\Omega_{e}\mid \sqrt{1-5p^{2}}$.
Its minimal frequency is $\omega_{up,min,SEx}$
which is located in area $\omega_{up,min,SEx}<\omega_{min,FEx}<\mid\Omega_{e}\mid$.
The small frequency wave exists in area $\omega\in (\omega_{l,up,min,SEx},\mid\Omega_{e}\mid \sqrt{1-5p^{2}})$,
while $\omega_{l,up,min,SEx}<\omega_{0}^{(3)}$.

We consider further modification of spectrum at the increase of temperature up to $\beta=1$.
It is represented in Fig. \ref{RHD2021ClLM Fig 02}.
General picture of spectrum is same for the large and small magnetic fields.
We have two wave solutions,
which are associated with slow and fast extraordinary waves.
If we compare this result with nonrelativistic limit
we see the disappearance of the fast magneto-sound wave.
If we compare it with the regime of small relativistic temperatures
we obtain that two wave solutions disappear.
The fast extraordinary wave starts at $\omega_{min,FEx}$:
$\omega_{min,FEx}<\omega_{0}^{(3)}<\omega_{0}^{(1)}$ if $b<1$,
and $\omega_{min,FEx}\ll\omega_{0}^{(3)}\ll\omega_{0}^{(1)}$ if $b>1$.
The spectrum of slow extraordinary wave starts at $\omega=0$
(let us remind that ions are motionless, otherwise this value changes).
The frequency increases with the increase of the wave vector,
it goes up to $\infty$ in accordance with $\omega\rightarrow kc$ at $k\rightarrow\infty$.
All features related to the cyclotron frequency disappear
since $\omega^{2}-\Omega_{e}^{2}(1-5p^{2})\neq 0$
due to $5p^{2}>1$ at $p^{2}<1$.

Further increase of temperature does not show any dramatic changes (see Fig. \ref{RHD2021ClLM Fig 03}).
The minimal frequency of the fast extraordinary wave becomes smaller
$\omega_{min,FEx}\ll\omega_{0}^{(3)}<\omega_{0}^{(1)}$ (for all $b$)
and the phase velocities of both waves becomes larger,
but these changes are rather small.

A relativistic hydrodynamics obtained from the kinetic model with the focus on the low frequency excitations for the magnetized plasmas is presented
in Refs. \cite{Hazeltine APJ 2002}, \cite{Mahajan PoP 2002} with the corresponding analysis of the spectrum.

\section{Conclusion}

Propagation of the electromagnetic waves parallel to the external magnetic field in the nonrelativistic plasmas is not affected by the thermal effects.
The relativistically hot magnetized plasmas has been considered.
It has been found that the thermal effects reveal themself at the propagation of the electromagnetic waves parallel to the external magnetic field.
Three regimes have been found.
\newline
1. Small relativistic temperatures ($T=0.1 m_{e}c^{2}$) and the small magnetic fields ($\mid\Omega_{e}\mid/\omega_{Le}=0.5$).
\newline
2. Small relativistic temperatures ($T=0.1 m_{e}c^{2}$) and the large magnetic fields ($\mid\Omega_{e}\mid/\omega_{Le}=2$).
\newline
3. Large relativistic temperatures $T\geq m_{e}c^{2}$ and arbitrary magnetic field.

Consider the first regime.
Three waves exist in the first regime like in the nonrelativistic limit.
Two high frequency waves extend area of their existence toward smaller frequencies.
However, the small frequency fast magneto-sound wave exists in the smaller interval of frequencies,
the maximal frequency is decreased by the relativistic thermal effects.

Consider the second regime.
Four waves exist in the second regime.
The interval of frequencies for the fast magneto-sound wave is described by the thermal effects from the high frequency side.
The interval of frequencies for the fast extraordinary wave is increased from the area of small frequencies.
In the nonrelativistic limit they areas of existence are separated by nonzero interval ($\mid\Omega_{e}\mid, \omega_{0}^{(1)}$).
The length of this interval changes and its position shifts in the area of small frequencies.
The slow extraordinary wave is splitted on two waves.
They areas of existence are also separated by an interval.
The low frequency part exists from nonzero minimal frequency to the thermally shifted cyclotron frequency.
The high frequency part exists from minimal frequency located above the thermally shifted cyclotron frequency up to infinity.

Consider the third regime.
The fast magneto-sound wave does not exist.
Fast and slow extraordinary wave extended the area of existence down to the rather small frequencies.
Fast extraordinary wave exists at frequencies above zero value.
Slow extraordinary wave exists at the frequency above minimal frequency.

The analysis is based on the model constructed of four hydrodynamic functions,
which reduces to two traditional functions: the concentration and the velocity field, in the nonrelativistic limit.
These functions are the concentration, the velocity field, the density of reverse relativistic factor, and
the current of the reverse relativistic factor.
These set of functions and equations of their evolution make up
the microscopically justified background for the hydrodynamic study of the collective phenomena in the relativistically hot plasmas.

\section{Acknowledgements}

Work is supported by the Russian Foundation for Basic Research (grant no. 20-02-00476).

\section{DATA AVAILABILITY}

Data sharing is not applicable to this article as no new data were
created or analyzed in this study, which is a purely theoretical one.

\appendix

\section{Appendix: Definitions of the hydrodynamic functions}

All microscopic definitions presented below are expressed via the same operator:
\begin{equation}\label{RHD2021ClLM formula for average}\langle ...\rangle\equiv\frac{1}{\Delta}\int_{\Delta}d\mbox{\boldmath $\xi$}
\sum_{i=1}^{N} ... \delta(\textbf{r}+\mbox{\boldmath $\xi$}-\textbf{r}_{i}(t)),\end{equation}
which is replaced by symbol $\langle ...\rangle$ to express equations in shorter form.

Once we use averaging (\ref{RHD2021ClLM formula for average}) in explicit to present the concentration of particles
$n(\textbf{r},t)$ in the arbitrary inertial frame \cite{Kuz'menkov 91}, \cite{Drofa TMP 96}, \cite{Andreev PIERS 2012}
\begin{equation}\label{RHD2021ClLM concentration definition} n(\textbf{r},t)=\frac{1}{\Delta}\int_{\Delta}d\mbox{\boldmath $\xi$}\sum_{i=1}^{N}\delta(\textbf{r}+\mbox{\boldmath $\xi$}-\textbf{r}_{i}(t)). \end{equation}
The definition of concentration (\ref{RHD2021ClLM concentration definition}) contains the integral operator.
This operator counts the number of particles in the vicinity of the point of space.
This method is suggested by Kuz'menkov L.S. a generalization of method suggested by Klimontovich Yu.L.
\cite{Klimontovich Plasma}, \cite{Klimontovich Dokl 62}, \cite{Weinberg Gr 72}.

The current of particles can be written in a short form: $\textbf{j}=\langle \textbf{v}_{i}(t)\rangle$ using definition (\ref{RHD2021ClLM formula for average}),
where $\textbf{v}_{i}(t)=d\textbf{r}_{i}(t)/dt$.
Presented definition of current $\textbf{j}$ gives total velocity of all particles being in the delta vicinity over the volume of the vicinity $\Delta$:
$\textbf{j}=\sum_{i\in\Delta}\textbf{v}_{i}(t)/\Delta$.
The average velocity can be introduced at the usage of the current $\textbf{v}=\textbf{j}/n$.
The average velocity is found as arithmetic mean.

Similar definitions are found for other functions \cite{Andreev 2021 05}:
$\Gamma=\langle \frac{1}{\gamma_{i}}\rangle$,
$t^{a}=\langle \frac{1}{\gamma_{i}}v_{i}^{a} \rangle -\Gamma v^{a}$,
$p^{ab}=\langle v_{i}^{a}v_{i}^{b} \rangle-n v^{a}v^{b}$,
$t^{ab}=\langle \frac{1}{\gamma_{i}}v_{i}^{a}v_{i}^{b} \rangle-\Gamma v^{a}v^{b}-t^{a}v^{b}- v^{a}t^{b}$,
for $M^{abcd}$ see equation (17) of Ref. \cite{Andreev 2021 05},
where $\gamma_{i}=1/\sqrt{1-\textbf{v}_{i}(t)^{2}/c^{2}}$.

\end{document}